\documentclass[12pt]{iopart}

\usepackage{graphicx}

\begin{document}

\title[Strange Quark Matter in Stars]{Strange Quark Matter in Stars: A
  General Overview}

\author{J\"urgen Schaffner--Bielich}

\address{Institut f\"ur Theoretische Physik/Astrophysik,
J. W. Goethe Universit\"at,
D--60054~Frankfurt am Main, Germany}

\ead{schaffner@astro.uni-frankfurt.de}

\begin{abstract} 
The physics of strange quark matter in the core of compact stars and recent
compact star data is reviewed. Emphasis is put on the possible existence
of a third family of strange quark stars.
\end{abstract}

\section{Introduction}

The study of the QCD phase diagram encompasses a plethora of phenomena.
High temperatures and low baryon number densities are realized in the
early universe and in the mini-bang in the laboratory, i.e.\ 
relativistic heavy--ion collisions. On the other hand, low temperatures
and high baryon number densities are the realm of compact stars, neutron
stars and quark stars. In the following, we address the physics of the
latter area of the QCD phase diagram, in the spirit of this meeting
focusing on the r\^ole of strangeness.

Neutron stars are the endpoint of stellar evolution of massive stars
with $M>8M_\odot$. They are compact, massive objects with typical radii
of about 10 km and masses of $(1-2) M_\odot$. Presently, more than 1500
pulsars, rotating neutron stars, are known. So far, the best determined
mass is the one of the Hulse-Taylor pulsar with $M=1.4411\pm 0.00035
M_\odot$ \cite{Thorsett99}. We will discuss more recent developments in
the determination of neutron star masses and radii in section
\ref{sec:nsdata}.

Let us explore the composition and structure of neutron stars starting
from the surface. There is a thin atmosphere up to a density of about
$10^4$ g/cm$^3$ of atoms, the outer crust has a density between $10^4$
g/cm$^3$ and $4\cdot 10^{11}$ g/cm$^3$ and consists of a lattice of
nuclei surrounded by free electrons. The inner crust starts beyond the
drip-line density with densities of $4\cdot 10^{11}$ g/cm$^3$ to
$10^{14}$ g/cm$^3$, where the lattice of nuclei is now immersed in a gas
of neutrons and electrons.  At even larger densities, one reaches the
core of the neutron star, a liquid of neutrons, protons, electrons and
other particles. The composition of the core of a neutron star is a
matter of intense debate.  Several more or less exotic phases have been
proposed to exist: pion condensation, kaon condensation, hyperon matter,
and finally quark matter including strange quarks (see e.g.\ 
\cite{Glen_book,Weber_book} for a discussion on the various phases). The
Bodmer--Witten scenario, that strange quark matter can be more stable
than ordinary nuclear matter, suggests that there exists a corresponding
class of compact stars, so called strange stars
\cite{Haensel86,Alcock86}, where the core consists of about equal
amounts of up, down and strange quarks. Only an outer crust can exist
for this kind of compact star, as unbound neutrons are converted to more
stable strange quark matter.  Let us start more conventional and look at
a free gas of particles of known hadrons, including pions, kaons and
hyperons. It turns out, that in $\beta$-equilibrium the first ``exotic''
hadron beside nucleons and leptons are hyperons: the $\Sigma^-$ appears
at $4n_0$, the $\Lambda$ at $8n_0$, where $n_0=0.15$ fm $^{-3}$ is
normal nuclear matter density \cite{Ambart60}. The maximum mass,
however, will be around $M=0.7M_\odot$ for a free gas of neutrons
\cite{OV39} substantially smaller than the required $M=1.44M_\odot$.
Hence, taking into account interactions (and interaction energy) is
crucial to describe neutron stars. Note, that this in contrast to the
case of white dwarfs, which can be well described by a free gas of
electrons and a lattice of nuclei.

\section{Neutron Star Modelling}
\label{sec:nsmodel}

\begin{figure}
\centerline{\includegraphics[width=0.40\textwidth]{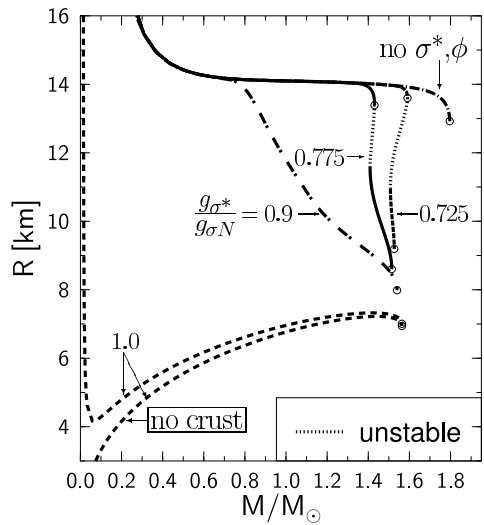}}
\caption{The mass--radius relation for neutron star matter including
  hyperons. For an increased attraction between hyperons, the
  mass-radius curve exhibits a new solution with similar maximum masses
  but smaller radii than ordinary neutron stars (solid lines). The
  dashed lines shows the case for absolutely stable hyperon matter
  (figure taken from \cite{Scha02}).}
\label{fig:mrhyp}
\end{figure}

Motivated by the results for a free gas of hadrons, implementing
hyperons into the composition of neutron star matter besides nucleons
and leptons seems to be a reasonable first step. Indeed, modern models,
of the equation of state for neutron stars confirm that the hyperons
$\Lambda$ and/or the $\Sigma^-$ appear first around $2n_0$, see
e.g.\ \cite{JSB04} and references therein.
Most notable, what one learns from these calculations is, that neutron
stars are most likely strange in the interior! The story about matter
with strangeness in neutron stars does not end here. Increasing the
strength of the basically unknown interaction between hyperons reveals a
new aspect of compact star physics: the possible existence of a new
class of family of compact stars besides white dwarfs and ordinary
neutron stars (see figure \ref{fig:mrhyp})! This new stable solution to the
Tolman--Oppenheimer--Volkoff (TOV) equations is characterised by an
increased compactness compared to ordinary neutron stars. The
mass-radius curve shows an instability at the onset of the phase
transition to hyperon matter which vanishes once a core of pure hyperon
matter is present \cite{Scha02}.
Increasing the attraction between hyperons even further, the mass-radius
curve changes its characteristics completely: hyperon matter gets
absolutely stable and corresponding hyperon stars can become arbitrarily
small in radii and mass. Typical radii even when including an (outer)
crust are in the range $R=7-8$ km. Such compact stars are generically
dubbed selfbound stars. Figure \ref{fig:mrhyp} shows the mass-radius
relation for the various cases. Note, that all these cases and phases
are described within one single Lagrangian!

In the following, let us summarise in some detail the properties of
selfbound stars and the third family. For neutron stars we note:
\begin{flushleft}
- they are bound by gravity, there is a finite pressure for all
energy density\\
- their mass--radius relation starts at large radii,\\
- the minimum neutron star mass is $M\approx 0.1M_\odot$ with
  $R\approx 200$ km\\
\end{flushleft}
while for selfbound stars:
\begin{flushleft}
- there is a vanishing pressure at a finite energy density\\
- the mass-radius relation starts at the origin (ignoring a possible
  outer crust)\\
- arbitrarily small masses and radii are possible.
\end{flushleft}
At first glance, the global properties of neutron stars and selfbound
stars, in particular those made out of absolutely stable strange quark
matter \cite{Haensel86,Alcock86}, are quite similar, like the maximum
mass and the corresponding radius, or the surface properties due to the
same outer crust. There are, however, unique features of strange stars,
which can have spectacular effects for astrophysical systems as e.g.:
compact stars with arbitrarily small masses and small radii
(dubbed golf balls in \cite{Hartle75}),
collapse of neutron stars to quark stars \cite{Ivan65},
explosive conversion of neutron stars to strange stars
  \cite{Olinto87},
generation of a secondary shock wave in supernova explosions
  \cite{derujula87,hatsuda87}, 
strange dwarfs: small and light white dwarfs with a strange
    star core \cite{GKW95},
super-Eddington luminosity from bare, hot strange stars
  \cite{Page02}. 

We stress, that the existence of strange stars hinges on the underlying
assumption that strange quark matter is more stable than ordinary
nuclear matter. On the other hand, the criteria for the existence of the
third family of compact stars is less stringent: there has to be a
strong first order phase transition in $\beta$-equilibrated matter
\cite{Gerlach68}. The new solution to the TOV equation is stable and
generates compact stars which are more compact than neutron stars. Note,
that {\em any} first order phase transition can possibly produce a third
family. Signals for a first-order phase transition in compact stars
and/or a third family in general proposed in the literature are e.g.:
delayed collapse of a proto-neutron star to a black hole
  \cite{Thorsson94},
spontaneous spin--up of pulsars \cite{Glen97},
generation of a secondary shock wave in supernova explosions
      (as proposed for strange stars),
specific waveforms of gravitational waves from colliding neutron
  stars \cite{Oechslin04},
collapse of a cold neutron star to the third family
\cite{Migdal79,Kaempfer81a} with possible emission of gravitational
  waves, $\gamma$-ray bursts, and neutrinos,
rising twins in the mass--radius diagram: compact stars with
  $M_1<M_2$ {\em and} $R_1<R_2$ (see e.g.\ \cite{Schertler00}),
delayed collapse of a cold neutron star and gamma-ray bursts
  \cite{BBDFL03}.

The third family of solution was considered before for hyperon matter
\cite{Heintzmann74} (so called hyperon stars), pion-condensed stars
\cite{Haensel80,Haensel82} and also for stars with a quark core
\cite{Kaempfer81,Kaempfer83}. Note, that K\"ampfer used polytropic
equation of states to derive conditions for the existence of a third
family for pion-condensed stars and stars with a quark core, so that his
results can be generalised in principle for an arbitrary exotic phase in
the core of compact stars.  A modern treatment investigating the
appearance of the third family with parametric equation of states can be
found in \cite{Macher04}. The third family solution was rediscovered
only recently. Its properties are now discussed in terms of modern
equation of states which take into account the Gibbs criteria for phase
transitions in compact stars
\cite{GK2000,Schertler00,Scha02,FPS01,Banik01,Mishustin03,Banik03}.
There is even a section devoted to the third family in the second edition
of Glendenning's textbook \cite{Glen_book}. Contrary to earlier findings
within Bose-condensed stars but consistent with the results of K\"ampfer
\cite{Kaempfer81,Kaempfer83}, the third family has similar masses
compared to the ordinary neutron star branch and constitutes a truly new
branch for compact stars. One should stress again, that the existence of
the third family of compact stars depends on a strong first order phase
transition of the equation of state (see e.g. the general discussion in
\cite{Gerlach68,Kaempfer81,Haensel82,Schertler00,Haensel03}). The underlying
micro-physical form of the phase transition is not relevant for the
global properties (mass--radius) of compact stars by solving the TOV
equations. So a third family has been found for a phase transition using
the MIT bag model for quark matter \cite{GK2000,Mishustin03}, massive
quasi-particles of quarks \cite{Schertler00}, hyperon matter
\cite{Scha02}, interacting quarks in perturbative QCD \cite{FPS01}, kaon
condensates \cite{Banik01}, and colour-superconducting quarks \cite{Banik03}.
The third family has been also studied for rotating compact stars in
\cite{Banik04,Bhatta04}.

In the following, we take a closer look at the QCD phase transition at
finite chemical potential and zero (or small) temperature focusing on
the chiral phase transition. For dense matter, there is a naive argument
why the chiral phase transition should happen after deconfinement.
Consider the counter-situation, assume that the hadrons are in the
chirally restored phase and hypothesise that the hadrons are massless
(which is not necessarily implied by chiral symmetry restoration!). Then
compare the pressure with that of free, massless quarks at the same
baryochemical potential and one finds that (see e.g.\cite{FPS01}):
$$
P_h(\mu_B) = N_h/(12\pi^2) \mu_B^4 = N_h/(12\pi^2) N_c^4 \mu_q^4 =
N_h/N_f N_c^3 P_q(\mu_B) \quad .
$$
Hence, the hadron pressure will be always larger than the quark
pressure and there will be no transition to quark matter possible!
Therefore, hadrons must transform to massive quarks first and then the
chiral phase transition to (nearly) massless quarks happens. Of course,
this line of arguments relies heavily on the tacit assumption that
interactions can be ignored!

The order and strength of the chiral phase transition actually
determines whether or not there is a third family of compact stars
\cite{FPS01,FPS02}. For a cross-over or a weak first-order phase transition,
there will be only one solution to the TOV equations for ordinary
neutron stars. For a strong first-order phase transition, however, there
will be two solutions, an additional one with smaller radii (and
probably smaller masses). These two scenarios outlined here actually
depends crucially on the {\em low-density} hadronic equation of state! If
the hadronic pressure rises strongly with the baryochemical potential,
the crossing with the quark pressure appears at large values of the
baryochemical potential and the change of the slope of the curves, which
is just the baryon number density, will be moderate. Hence, the
transition will be at most weakly first order. On the other hand, if the
hadron pressure increases weakly with the baryochemical potential, the
phase transition will happen at smaller baryochemical potential and the
change of the slope changes drastically. In that case, the transition
will be strongly first order. Unfortunately, the low-density equation of
state of hadronic matter is still not well determined. The only thing
one can state is that a strong first order phase transition is not ruled
out by calculations of asymmetric nuclear matter up to $2n_0$
\cite{Akmal98}: the hadronic pressure increases slowly and is still
considerably smaller compared to that of free quarks, i.e. only about
4\% of $P_q$ at $n_0$ \cite{FPS02}.

Using perturbative QCD as a model for dense QCD, the matching to the
hadronic equation of state indeed can produce a third family of compact
stars with a large quark core for reasonable transition densities
\cite{FPS02}. Note that the maximum mass configuration of ordinary
neutron stars contains then also a mixed phase of hadrons (or massive
quarks) and chirally restored quarks in the core.  The difference to the
quark star solution of the third family is that their core consists of
pure (chirally restored) quark matter surrounded by layers of a mixed
phase and a hadronic (chirally broken) phase.

Recently, the study of quark stars has been refined by including effects
from colour-superconductivity (see
\cite{Alford03,Lugones03,Baldo03,Banik03,Blaschke03,Shovkovy03,Grigorian04,Ruster04,Buballa04,Drago04,Alford04}).
Here we just quote some of the recent findings and refer the interested
reader to the contribution of Shovkovy, R\"uster and Rischke for more
details \cite{Shovkovy04}. It was shown by R\"uster and Rischke, that an
increased interaction between quarks results actually in an increased
mass for quark stars \cite{Ruster04}.  On the other hand, matching of an
hadronic equation of state with a quark matter equation of state, which
incorporates features of perturbative QCD and colour-superconductivity,
can give mass-radius relations which are indistinguishable from ordinary
purely hadronic stars \cite{Alford04}.  Most recently, the phase diagram
of colour-superconducting matter in $\beta$ equilibrium has been studied
at finite temperature \cite{Ruster04pd,Fuku04}. Besides the more
standard two-flavour colour-superconducting (2SC) phase and the
colour-flavour-locked (CFL) phase a rich variety of new phases appears,
like the gapless CFL phase and the metallic CFL phase with exotic
properties. The ungapped normal quark phase is present only at quite large
temperatures of more than 50 MeV. If quark matter is formed during
cool-down of a hot proto-neutron star, the matter in the core will pass
through these phases, experiencing various phase transitions.  Hence,
those exotic phases are important for the physics of newly born
proto-neutron stars!

\section{Neutron Star Data}
\label{sec:nsdata}

The recent years have shown many surprises in the observations of
compact stars. There are trivial constraints on the mass and radius of a
compact star. Besides the Schwarzschild radius of $R_s=2GM$ a compact
star has to be larger than about $3GM$ for causal equations of state
\cite{Glen2000}. The radius for a $1.4M_\odot$ star has to be smaller
than 15.5 km to be compatible with the fastest rotating pulsar PSR
1937+21 with a spin frequency of 641 Hz \cite{Glen92limit,LP04}.

The x-ray pulsar Vela X--1 has a measured mass of $M=1.88\pm0.13M_\odot$
\cite{Quaintrell03}. If confirmed, the measurement would constitute a
new lower limit to the maximum mass of neutron stars. The mass
measurement for the high-mass x-ray binary U1700-37 with $M=2.44\pm0.27$
is well above $M(2\sigma)>2M_\odot$ but it can not be ruled out that
this is a black hole and not a neutron star \cite{Clark02}. An
interesting candidate for a heavy neutron star has been reported in
\cite{Nice04,Nice04b}: the pulsar PSR J0751+1807 with a white dwarf
companion has a measured mass limit of $M>1.6M_\odot$ (within
$2\sigma$)! The error bars are still quite large but will be reduced as
more data is collected.

A highly debated compact star is the isolated neutron star RX
J1856.5-3754 which is the closest one known. The x-ray data shows a
perfect black-body spectra \cite{Drake02}. However, no spectral lines
are detected and the optical flux is not compatible with the
extrapolated simple black-body formula.  All classical neutron star
atmospheres are basically ruled out (hydrogen as well as heavy element
atmospheres) \cite{Burwitz03}. An alternative description with a
two-component black-body fit comes to a surprising conclusion: the soft
temperature must be small so that the x-ray part of the spectra is not
spoiled. This on the other hand implies a lower limit for the emitting
radius to get the optical flux right: $R_\infty > 16.5 \mbox{ km}
(d/117\mbox{ pc})$, where $R_\infty$ is the radius measured by an
observer at infinite distance \cite{Truemper03}. As $R_\infty=
R/\sqrt{1-2GM/R}$, allowed masses and radii are very large and nearly
every equation of state on the market would be ruled out, except for
extremely hard equations of state with no phase transition at all to any
exotic matter! We stress, however, that this will not rule out a
possible third family of compact stars. It might well be that this
isolated neutron star has no exotic core, but still there might be
compact (quark) stars with a considerably smaller radius (and maybe
smaller masses).

There is a promise from x-ray bursters to get another handle on the
compactness of a neutron star by measuring spectral lines. In a binary
system with a neutron star, the neutron star is accreting material from
its companion which ignites a thermonuclear explosion on the surface of the
neutron star. Measured spectral lines, if identified correctly, are
redshifted and give a direct measure of the $M/R$-ratio. For EXO
0748-676 \cite{Cottam02} a value of $z=0.35$ was extracted from three
different spectral lines which gives $M/M_\odot=1.5(R/10 \mbox{ km})$. A
more independent way to explore the gravitational potential around
compact stars is to measure the spectral profile of emitted spectral
lines which is modified from the space-time warpage \cite{Strohmayer04}.
A constraint of $9.5 \mbox{ km} < R < 15 \mbox{ km}$ was found for EXO
0748-676 \cite{Villa04}, the same object studied above. Using $z=0.35$
from \cite{Cottam02} implies a mass range of $1.5M_\odot < M < 2.3 M_\odot$
\cite{Villa04}.

In the future, the observations of compact stars will be revolutionised
by the Square Kilometre Array (SKA). The SKA is an international project
to built a receiving surface of one million square kilometres. The
potential to discover are 10,000 to 20,000 new pulsars, more than 1,000
millisecond pulsars and at least 100 compact relativistic binaries
\cite{Kramer03}! These future measurements will probe the equation of
state at extreme limits! In addition, the SKA can be used as a gigantic
cosmic gravitational wave detector by using pulsars as clocks. The
design and location is not finalised yet; maybe the host country of this
conference, South Africa, being one of the candidates, will be
successful! In any case, the future is bright for peering into the heart
of compact stars!

\section*{References}

\bibliographystyle{revtex}
\bibliography{all,literat}

\end{document}